\documentclass[aps,prl,twocolumn,showpacs,preprintnumbers,amsmath,amssymb,floatfix]{revtex4}

\usepackage{graphicx}
\usepackage{dcolumn}% Align table columns on decimal point
\usepackage{bm}% bold math

\begin{document}

\title{Many-body excitations in tunneling current spectra of a few-electron quantum dot
%Tunneling current magnetospectroscopy in a few-electron quantum dot
}

\author{D. V. Melnikov,  T. Fujisawa$^{\dagger}$, D. G. Austing$^{\dagger\dagger}$, S. Tarucha$^{\dagger\dagger\dagger}$, and J.-P. Leburton}

\affiliation{
Beckman Institute for Advanced Science and Technology \& Department of Electrical and Computer Engineering, University of Illinois at Urbana-Champaign, 405 North Mathews Avenue, Urbana, IL 61801, USA\\
$^{\dagger}$NTT Basic Research Laboratories, NTT Corporation, Atsugi-shi, Kanagawa 243-0198, and\\ Tokyo Institute of Technology, 2-12-1 Okayama, Meguro-ku, Tokyo 152-8551, Japan\\ $^{\dagger\dagger}$Institute of Microstructural Sciences M50, National Research Council of Canada, Montreal Road, Ottawa, Ontario K1A 0R6, Canada\\
$^{\dagger\dagger\dagger}$ICORP Spin Information Project, JST, Atsugi-shi, Kanagawa 243-0198, and\\ 
Department of Applied Physics, University of Tokyo, Hongo, Bunkyo-ku, Tokyo 113-8656, Japan
}

\date{\today}

\begin{abstract}

Inherent asymmetry in the tunneling barriers of few-electron quantum dots induces intrinsically different tunneling currents for forward and reverse source-drain biases in the non-linear transport regime. Here we show that in addition to spin selection rules, overlap matrix elements between many-body states are crucial for the correct description of tunneling transmission through quantum dots at large magnetic fields. Signatures of excited $(N-1)$-electron states in the transport process through the $N$-electron system are clearly identified in the measured transconductances. Our analysis clearly confirms the validity of single-electron quantum transport theory in quantum dots.

%the computed current spectra are in good agreement with the experimental data.

%The non-linear tunneling current through a well defined few-electron vertical quantum dot in a magnetic field is investigated. It is shown that the asymmetry in the tunneling barriers and spin selection rules dominate the current spectra, and the overlap matrix elements are crucial for the correct description of the tunneling through the ground and excited states at large magnetic fields. On the basis of the calculations, features of the experimental differential transconductance due to the involvement of the excited $N$-electron states in the transport process through the $(N+1)$-electron system are also identified. Overall, the computed current spectra are found to be in a very good agreement with the experimental data.

\end{abstract}


\maketitle

Lithographically defined quantum dots (QDs) are nanoscale systems in which electrons are constrained by quantum confinement \cite{Reimann,RPP} created by hetero-structure barriers and a spatially separated dopant charge distribution around the dot (in the leads) modulated by external gate voltages. The resulting potential in the plane of the dot has an approximately parabolic dependence with distance from the center of the QD \cite{Matagne}. %The rich electronic properties of these QDs can be readily accessed by measuring the transport characteristics in both the linear and non-linear transport regime.

Usually the QDs electronic structure is probed by single-electron transport spectroscopy where electrons tunnel through the double barrier structure under the influence of a source-drain $V_{SD}$ \cite{Kouw,Lateral}. In this approach, the non-linear regime, for which $V_{SD}$ is of the order of the average energy separation between many-body states, gives access to the QD excited states. However, this technique has its shortcomings since the numerous features observed in the current spectra \cite{Kouw} cannot be solely explained in terms of the QD energy spectrum.

Indeed, during single-electron tunneling processes, the QD electronic configuration fluctuates between $N$- and $(N-1)$-electron states, so that carrier transmission depends on the detailed occupation of the many-body energy states \cite{Devoret,Ashoori}. Hence, in a structure with asymmetric barriers, electrons injected in the QD through the thick (forward bias) rather than the thin (reverse bias) emitter barrier achieve different electron configurations, as in the former case the dot will be in the $(N-1)$-electron state most of the time (because of the thin collector barrier allowing easy escape of electrons from the QD), while in the latter case it will be predominantly occupied by $N$ electrons \cite{Averin}.

In this work we show that in a few-electron QD with asymmetric hetero-barriers distinct tunneling current spectra emerge for forward and reverse $V_{SD}$ as a result of different non-equilibrium QD state occupations \cite{Averin,Been,Kinaret}. Moreover, our analysis shows that the consideration of quantum-mechanical overlaps between many-body states \cite{Kinaret,Pfan} and spin selectivity of the tunneling are essential for understanding the non-linear transport characteristics in QDs.

Our experimental QD structure has a layered profile in the vertical direction, and is similar to structures used earlier \cite{RPP, Kouw}, {\it i. e.,} the two Al$_{0.22}$Ga$_{0.78}$As barriers and In$_{0.05}$Ga$_{0.95}$As quantum well are nominally 7.5 (thinner barrier with calculated tunneling rate $\gamma_{L}\sim 3.6$ ns$^{-1}$), 9.0 (thicker barrier with calculated tunneling rate $\gamma_{R}\sim 0.5$ ns$^{-1}$) and 12 nm thick, respectively. The QD formed in the quantum well region is controlled by a voltage $V_G$ applied to a single gate wrapped around the base of the cylindrical mesa with a diameter of 0.54 $\mu$m \cite{RPP}. The current-voltage characteristics are measured at a temperature $\sim 0.1$ K. Since the dot-lead coupling is relatively weak, the maximum current measured at 1.5 mV is $\sim 50$ pA (conductance $\partial I/\partial V_{SD}$ is less than $10^{-3}e^2/h$).

\begin{figure*}[tbp]
\includegraphics[width=8.6cm]{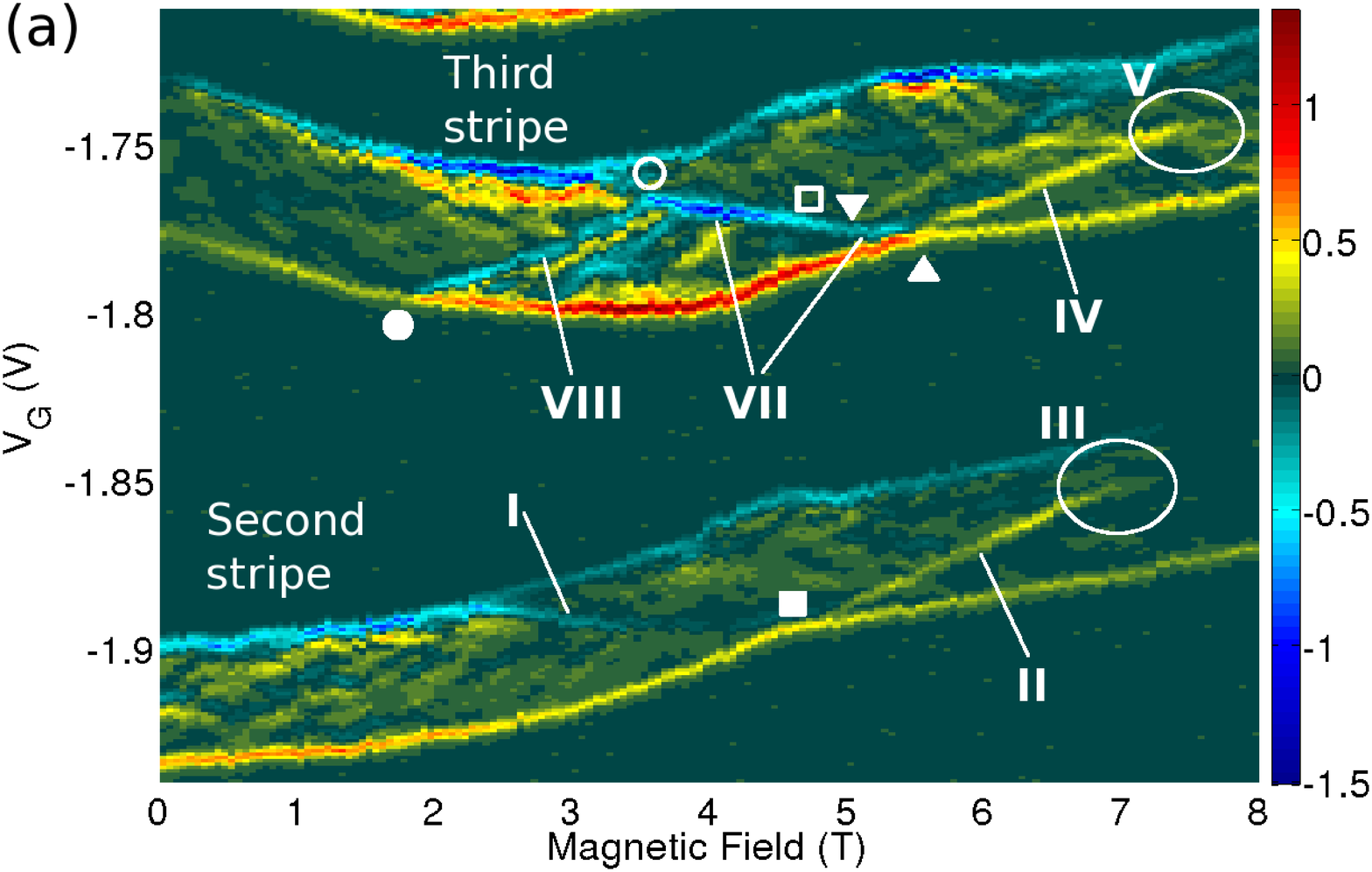}\includegraphics[width=8.6cm]{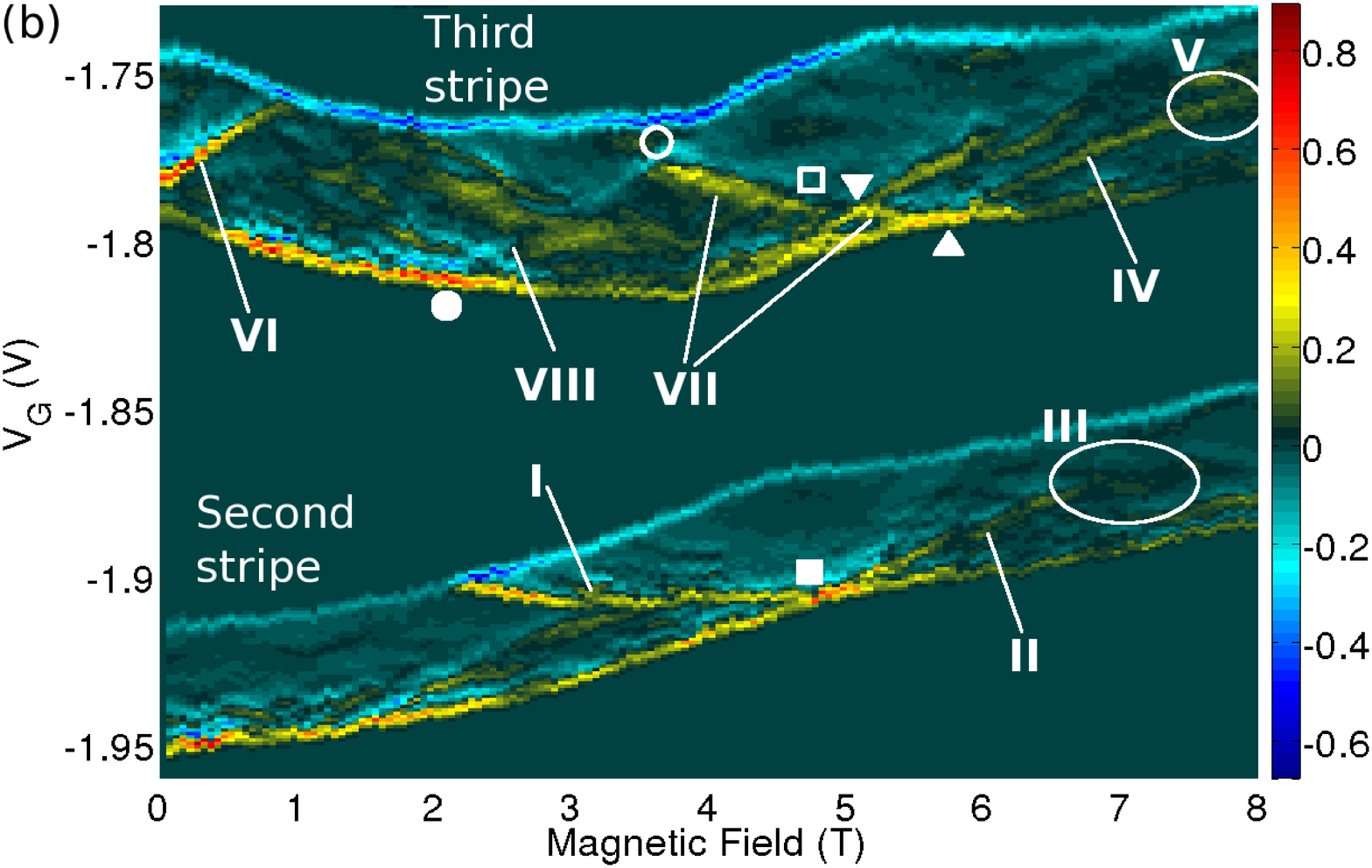}
\includegraphics[width=8.6cm]{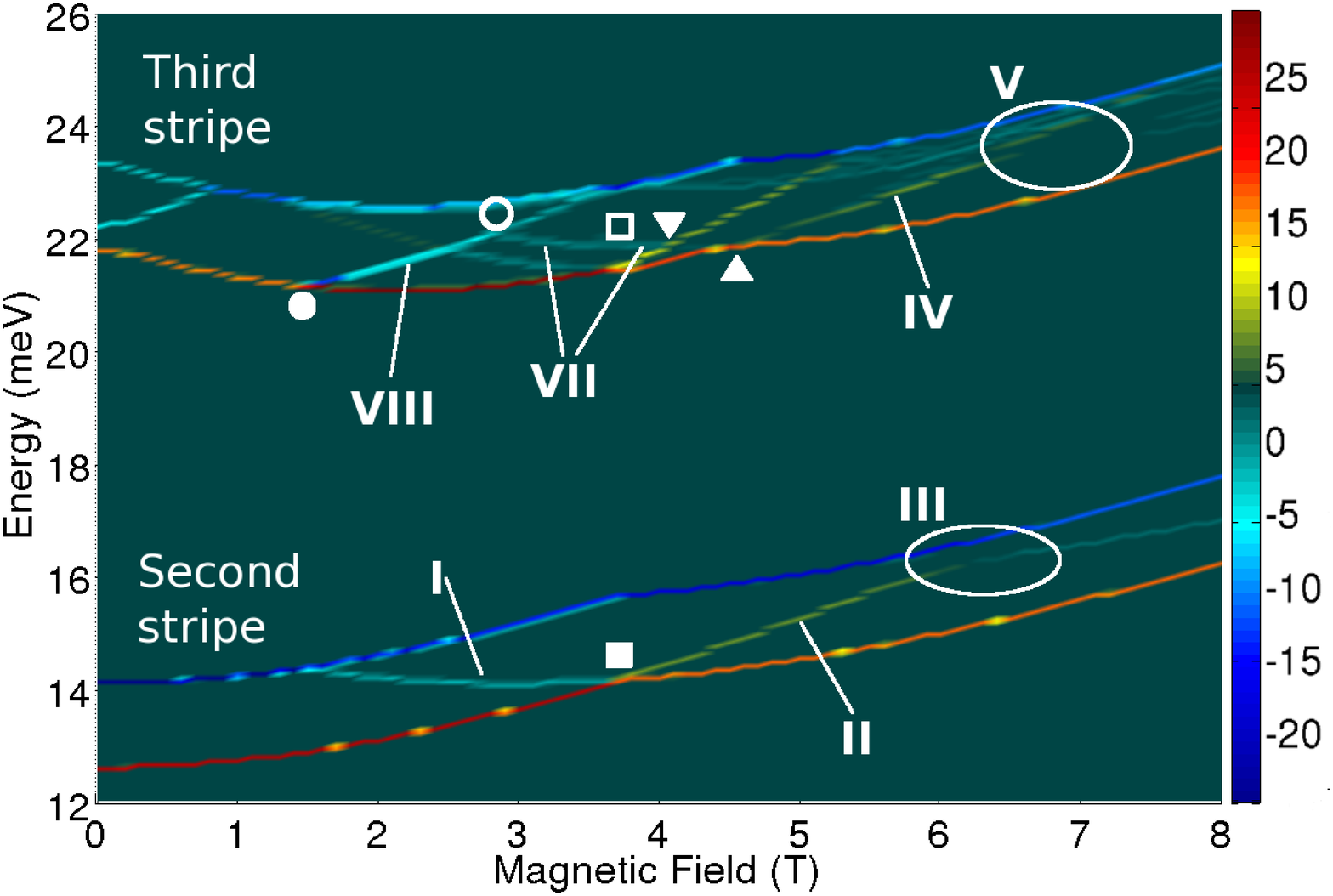}\includegraphics[width=8.6cm]{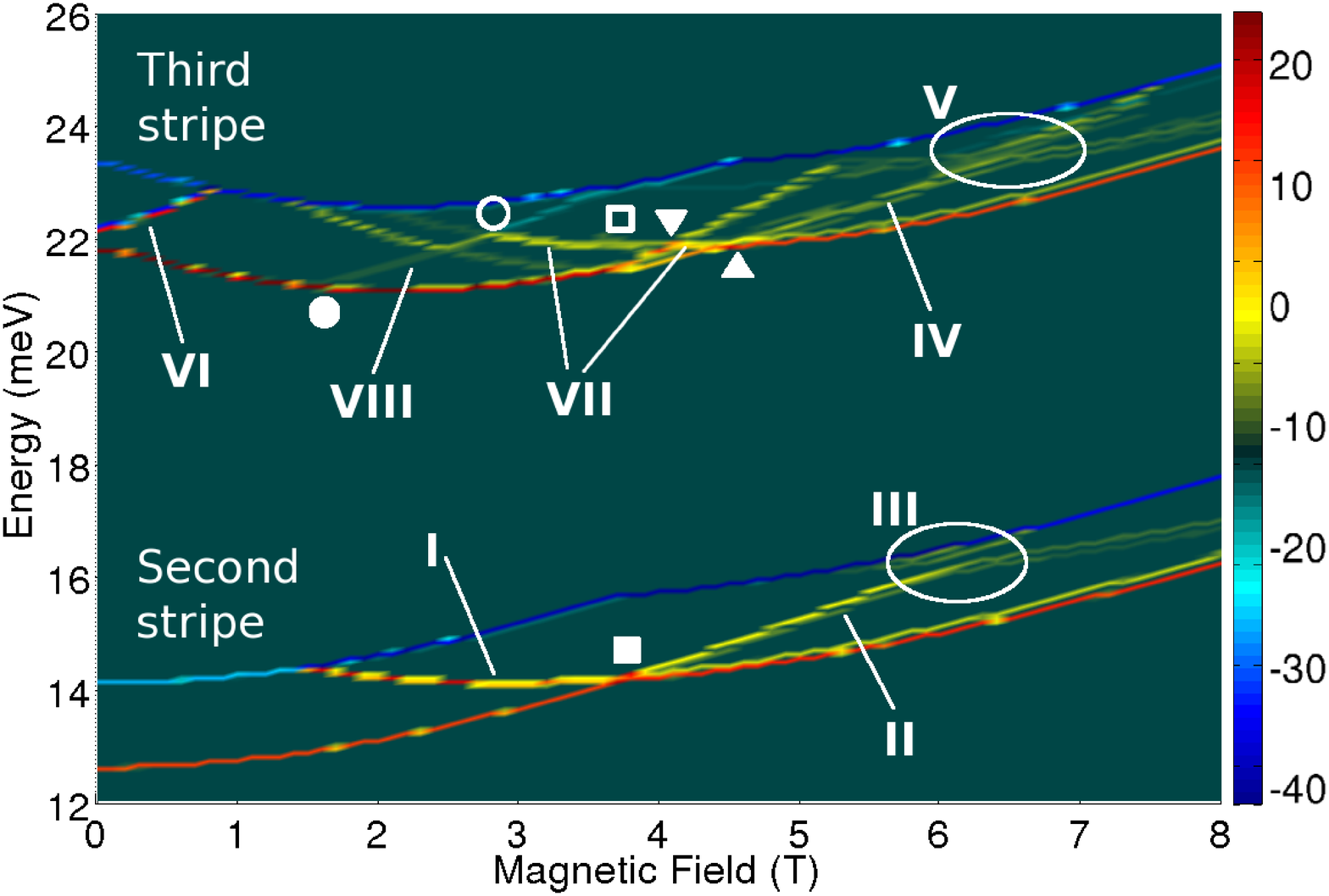}
\caption{Measured (top row) and computed (bottom row) differential transconductance plots for (a) the forward ($V_{SD}=1.5$ mV) and (b) reverse ($V_{SD}=-1.5$ mV) source-drain bias. To obtain $\partial I/\partial V_G$ in real units, the measured and computed data should be multiplied by 5 pA/mV and 80 pA/meV, respectively.
%Lower (higher) stripe corresponds to second (third) current stripe and provides information about $N=2$ ($N=3$) electron system. 
Symbols and features (regions) I to VIII are discussed in the text.
}
\label{fig:1}
\end{figure*}

Contour plots of the measured differential transconductance $\partial I/\partial V_G$ as a function of the gate voltage $V_G$ and magnetic field are shown in Fig. \ref{fig:1}, top row, for forward [$V_{SD}=1.5$ mV, Fig. \ref{fig:1}(a)] and reverse [$V_{SD}=-1.5$ mV, Fig. \ref{fig:1}(b)] source-drain biases where we 
%define forward (reverse) bias such that electrons are injected into the dot through the thinner (thicker) emitter barrier and ejected out of the dot through the thicker (thinner) collector barrier and 
take $I$ as the absolute value of the current. The finite transport window gives rise to non-zero current over a finite range of $V_G$ thus forming current stripes. Inside each stripe, there are somewhat weak random-looking features that are attributed to fluctuations in the density of states in the leads \cite{Kouw} as well as a number of more pronounced features due to excitations in the many-electron system. 

From Fig. \ref{fig:1}, top row, one can see that for $N=2$, aside from the second stripe's lower and upper edges, there is only one current step within the 1.5 mV transport window (feature I) at low magnetic field, merging with the lower edge of the stripe at about 5 T ({\footnotesize $\blacksquare$}), followed at higher field by a current step (feature II) that disappears at $\sim 7$ T (region III).

\begin{figure}[ptb]
\includegraphics[width=8.6cm]{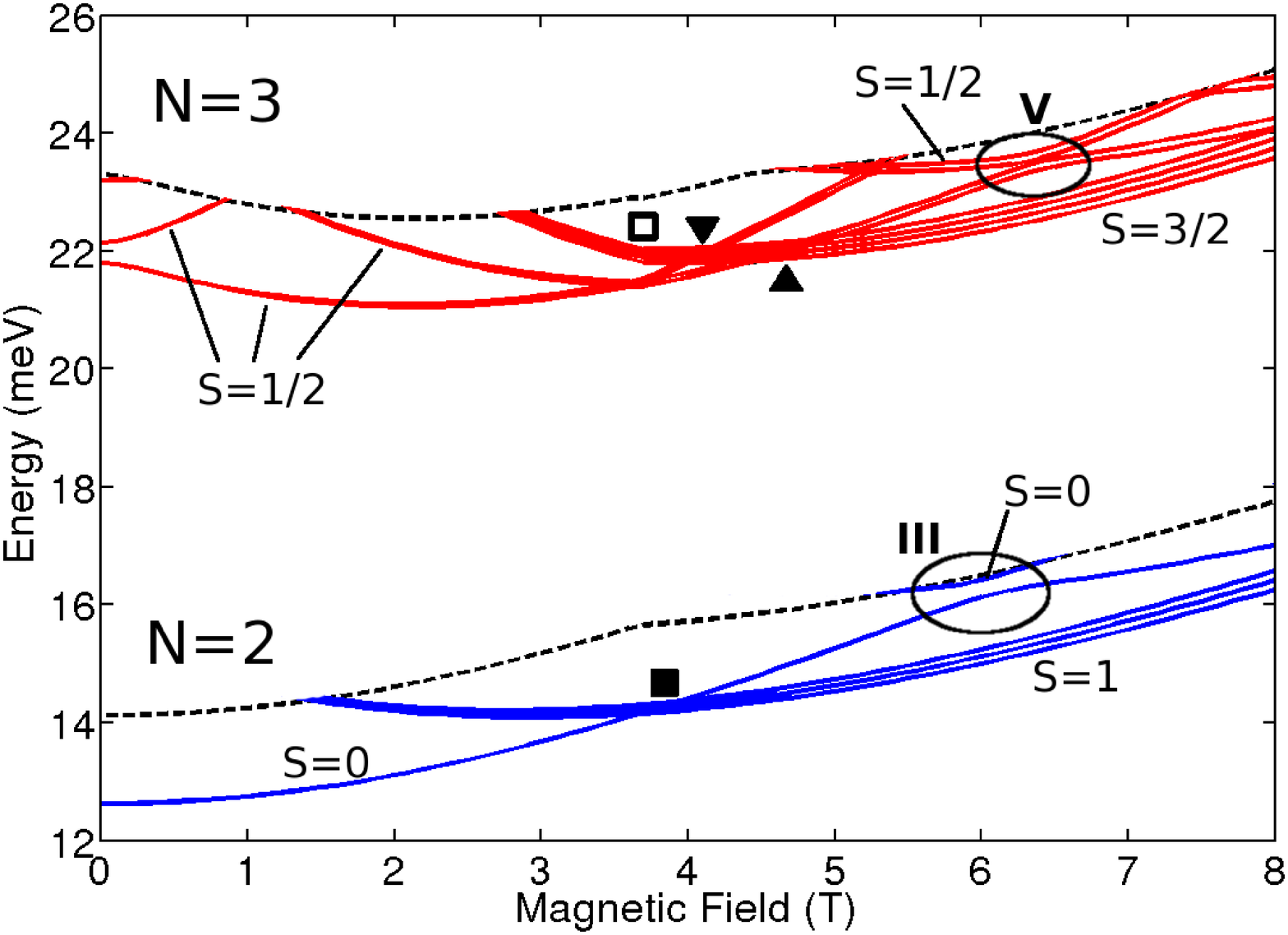}
\caption{Energy spectrum (electro-chemical potentials) for $N=2,~3$ computed with respect to the $(N-1)$-electron ground state energy within a 1.5 meV window. Dashed lines show the upper boundaries of the transport windows. Zeeman splitting of the energy levels manifests itself at higher magnetic fields as multiple closely separated lines unresolved in the experiment. Symbols and features (regions) are the same as in Fig.~\ref{fig:1}.
}
\label{fig:2}
\end{figure}

In order to gain insight in the current spectra, we compute the energy spectrum of the $N$-electron QD by numerically diagonalizing the Hamiltonian of the system $\hat{H}$~\cite{dmm3}:
\begin{equation}
\hat{H}=\sum_{i=1}^N\hat{h_i}+\sum_{i<j}\frac{e^2}{\epsilon{\bf |r_i-r_j|}},
\label{eq:ED}
\end{equation}
where the single-particle Hamiltonians $\hat{h_i}$ are ($\hbar=1$)
\begin{equation}
\hat{h_i}=-\frac{1}{2m^*}\left(\nabla_i-\frac{ie}{c}{\bf 
A}_i\right)^2+V_{c}({\bf r}_i)\pm\frac{1}{2}g\mu_BB
\label{eq:SPH}
\end{equation}
and the second term in Eq. (\ref{eq:ED}) describes the Coulomb interaction among the electrons. $m^*=0.061m_0$ and $\epsilon=13.0$ are respectively the conduction band effective mass and dielectric constant in In$_{0.05}$Ga$_{0.95}$As, and ${\bf A}=(B/2)(-y,x,0)$ is the vector potential in the symmetric guage for a magnetic field $B$ oriented perpendicular to the QD plane. The last term in $\hat{h_i}$ accounts for Zeeman splitting for which we set $g=-0.44$. Due to lithographic and natural imperfections \cite{RPP}, the confinement potential $V_{c}({\bf r})$ of the studied QD is known to be slightly elliptic, {\it i.e.}, $V_{c}({\bf r})=(1/2)m^*\left [\omega_x^2x^2+\omega_y^2y^2\right ]$ with the confinement energies in the $x$- and $y$-directions being $\hbar\omega_x=5.3$ and $\hbar\omega_y=5.65$ meV, respectively. Fig. \ref{fig:2} shows the QD energy spectra for $N=2,~3$ electrons.

By comparing with the experimental spectrum in Fig.~\ref{fig:1}, top row, one can see that features I and II correspond respectively to the excited triplet and singlet states that cross at $\sim 4$ T \cite{Kouw, footnote1}. Since with increasing number of electrons the separation between energy levels generally decreases, the energy window for $N=3$ in Fig. \ref{fig:2} contain more features whose evolution with magnetic field is similar to features in the $N=2$ stripe. Most notably the current step (feature IV) in Fig. \ref{fig:1} visible only up to $~\sim 7$ T (region V) is due to the lowest excited doublet. Strong features in the $N=3$ stripe between 4 -- 6 T (Fig. \ref{fig:1}) can be assigned to the lowest quartet state crossing with the two lowest doublet states [$\blacktriangle$ and $\blacktriangledown$] which are clearly distinguishable despite their proximity. The current step at feature VII between $\square$ (the magnetic field at which the singlet-triplet transition {\footnotesize $\blacksquare$} occurs for $N=2$) and $\blacktriangledown$ due to the ground state $N=3$ quartet and the ground state $N=2$ triplet \cite{Section, Palacios} is visible in both Fig. \ref{fig:1}(a) and (b). The first excited spin doublet state is also clearly seen in Fig. \ref{fig:1}(b) below 1 T (feature VI). Note that the splitting between this excited state and the $N=3$ ground state at 0 T due to the slight asymmetry of the QD is used to obtain appropriate values of $\omega_x$ and $\omega_y$ in $V_{c}({\bf r})$.

However, despite these agreements, there are other features in the experimental data of Fig.~\ref{fig:1} which cannot be account for by the energy spectrum alone. For example, the current steps due to the lowest excited $N=2$ singlet and the lowest excited $N=3$ doublet "disappear" altogether above $\sim 7$ T (beyond regions III and V), even though the spectra calculations show these energy states to be present. Also, the distinctive current steps between {\LARGE $\circ$} and $\square$ (feature VII) and {\LARGE $\bullet$} and {\LARGE $\circ$} (feature VIII) in Fig. \ref{fig:1} do not have any obvious counterparts in the calculated energy spectrum of Fig. \ref{fig:2}. More importantly we can clearly see the effect of the inherent asymmetry of the double-barrier hetero-structure on the measured $\partial I/\partial V_G$ as different steps in the current at features I, II and VI in Figs. \ref{fig:1}(a) and (b).

With the eigenspectrum of the $N$-electron system [Eqs. (\ref{eq:ED},~\ref{eq:SPH})], we perform explicit calculations of the (sequential) current $I$ through the QD \cite{Kinaret}:
\begin{equation}
I=-e\gamma\sum_{\alpha\beta}\Gamma_{\alpha\beta}\left[P_{\alpha}(N)+P_{\beta}(N-1)\right]\left[f_L-f_R\right ],
\label{eq:current}
\end{equation}
where $\gamma=\gamma_L\gamma_R/(\gamma_R+\gamma_L)\sim 0.45$ ns$^{-1}$ is the effective "bare" electron coupling between the QD and the leads~\cite{Brouwer}, 
%corresponding to the current $e\gamma\sim 80$ pA, 
$f_{L(R)}=f_{L(R)}\left [\Delta E_{\alpha\beta}-\mu_{L(R)}\right ]$ are the Fermi functions that determine the energy level occupation for electrons tunneling in to (out of) the QD from (to) the left (right) lead $L$ ($R$) with chemical potential $\mu_L$ ($\mu_R$), and $\Delta E_{\alpha\beta}=E_{\alpha}(N)-E_{\beta}(N-1)$ is the difference between the $\alpha$-th $N$-electron and $\beta$-th $(N-1)$-electron energies (the electro-chemical potential). $\Gamma_{\alpha\beta}$ is the overlap matrix element given by
\begin{equation}
\Gamma_{\alpha\beta}=\sum_{i}\left|\left<\Psi_{\alpha}(N)\left|a_{i}^{\dagger}\right|\Psi_{\beta}(N-1)\right>\right|^2
\label{eq:overlap}
\end{equation}
where $\Psi_{\alpha}(N)$ and $a^{\dagger}_i$ are the wave function of the $\alpha$-th $N$-electron state and the operator creating an electron in the QD in the $i$-th single-particle state, respectively. The non-equilibrium occupation factors $P_{\alpha}(N)$ in Eq. (\ref{eq:current}) can be determined from the steady state solution of the coupled master (or rate) equations \cite{Averin, Been, Ralph}. In our calculations a large number of eigenstates (up to 24 for each $N$) is included in the solution of the master equations to describe degeneracy effects in $\Delta E_{\alpha\beta}$.

The computed differential transconductance for $V_{SD}=\pm1.5$ mV is presented in Fig. \ref{fig:1}, bottom row. The maximum calculated current is $\sim 180$ pA in good agreement with the measured current, given mono-layer fluctuations in the QD structure \cite{footnote2}. The plots reproduce most of observed features shown in the top panels. For each stripe the current starts to flow (stops) when the ground state electro-chemical potential of the QD becomes lower than the chemical potential of the source $\mu_L$ (drain $\mu_R$), as expected. However, within the transport window, the current behavior is not trivial, and depends strongly on the structure asymmetry (and sign of source-drain bias) and the magnetic field.

The non-trivial behavior of the current within the second stripe ($N=2$) in Fig. \ref{fig:1} can be understood by noting that in our strongly asymetric structure ($\gamma_L/\gamma_R\sim 7$) for $V_{SD}>0$, the QD is occupied most of the time by two electrons, {\it i.e.}, $P_{\alpha}(N=2)\approx 1/M$, where $M$ is the number of two-electron states in the transport window.
%, and $P_{\alpha}(N=1)\rightarrow 0$. 
On the other hand, for negative $V_{SD}$, $P_{\alpha}(N=2)\approx\gamma_R/(2\gamma_L)\ll 1$, independent of the number of the states.
% and $P_{\alpha}(N=1)\approx 1/2$.
As spin-polarized triplet states for which $|S_z(N)-S_z(N-1)|>1/2$ is forbidden (the overlap matrix element for the spin-polarized triplet state $S_z(N=2)=\pm 1$ is either zero or $\approx 1$ depending on the spin of the $N=1$ system $S_z(N=1)=\pm 1/2$), the current through the $S=1$ state is always partially blocked. This means that with a thick collector barrier ($V_{SD}>0$, feature I), the overall current decreases compared to the current at the onset of the stripe \cite{Palacios}. However, in the case of a thin collector barrier ($V_{SD}<0$), the current rises when the triplet enters the transport window because in this case the $P_{\alpha}(N=2)$'s remain unchanged.
At high magnetic fields, beyond the singlet-triplet transition ({\footnotesize $\blacksquare$}), the triplet becomes the ground state at the lower edge of the $N=2$ stripe. At the same time, the current increases for both positive and negative biases when the excited singlet state (feature II) enters the transport window since no current blocking is possible for this state.
At $B \sim 6$ T (region III), the crossing between the two singlet states with different angular momenta \cite{Wagner} leads to a decrease of the overlap integral for the lower state from $\approx 0.9$ to $\approx 0.25$ resulting in a reduced current step for this state above 7 T, and thus explaining its "fading" beyond region III in Fig. \ref{fig:1}. Note a crossing between the two excited doublet states in the third stripes $\sim 7$ T (region V) has a similar effect on the current in that region.

Features VII and VIII in the third current stripe can be understood in terms of single-electron tunneling \cite{Kouw1} as follows. At the lower edge of the current stripe, when an extra electron is added to the two-electron system in the ground state, the three-electron ground state is formed. Within the stripe, however, if the transport window is sufficiently large, {\it any} of the three electrons in the QD can tunnel out leaving the two-electron system in either the singlet or triplet state with a probability given by the value of the overlap integral (\ref{eq:overlap}). As another electron can be injected into the QD in any of these two-electron states, this process effectively gives rise to additional transport channels within the bias window. An example is feature VII between {\LARGE $\circ$} and $\square$ which is attributed to electron addition to the lowest $N=2$ excited triplet forming the $N=3$ excited quartet \cite{Section} and along which, in forward bias, as for feature I in Fig. \ref{fig:1}(a), the current also decreases (the quartet-singlet transition is forbidden by the spin blockade condition \cite{Palacios}).
Feature VIII occurs when the electro-chemical potentials $\Delta E_{\alpha\beta}$ associated with the lowest $N=3$ doublet and the lowest $N=2$ triplet drop below $\mu_R$. The decrease in the number of channels leads to a strong reduction in current in Fig. \ref{fig:1}(a) and a slight increase in the current in Fig. \ref{fig:1}(b). The latter happens because of the redistribution of electrons among available channels when the partially spin blocked transport channels (due to the involvement of the triplet state, cf. feature I) become energetically inaccessible \cite{footnote3}. This is, however, inverse to the feature I in Fig. \ref{fig:1}(a) where the appearence of the corresponding channels in the transport window lead to the suppression of the current.
%The opposite current behavior occurs at feature I in Fig. \ref{fig:1}(a) where the appearence of the corresponding channels in the transport window lead to clear suppression of the current.

In general, as the number of $\Delta E_{\alpha\beta}$ with values between $\mu_L$ and $\mu_R$ can be very large, these effects can drastically influence the QD transport properties for any $N$. However, for small $V_{SD}$ quantum mechanical overlap (\ref{eq:overlap}) forbids transport via the majority of the state pairs in the transport window ($\Gamma_{\alpha\beta}\approx0$), though our simulations do show that with increasing $V_{SD}$ the current spectra rapidly become more complicated. For $N=2$ this is expected to occur for biases exceeding 2 mV in the magnetic field range of interest when the first excited one-electron state pops into the transport window. Also, at high magnetic fields, the large density of states and the number of the transport mechanisms involved will result in smaller changes in current for each individual transport channel [due to the relatively small variations among $P_{\alpha}(N)$], leading to overall decrease in visibility of features in the current stripes.

In the above analysis, we assumed no momentum or spin relaxation \cite{Hawrylak}. Since the slowest tunneling time in our QD is considerably shorter than the spin relaxation time which  has been measured to be $\sim 100~\mu$s in vertical QDs \cite{JCM}, the neglect of spin relaxation should not affect noticeably the current, {\it i.e.}, electrons will tunnel out of the QD long before the relaxation causes a change in the level occupation. Also, assuming the momentum relaxation time $\sim 10$ ns measured for $N=1$ \cite{JCM} remains valid for $N>1$, this relaxation mechanism may lead to a partial suppression of the current steps for some of the excited states for $V_{SD}>0$. For example, this could explain why the first excited $N=3$ spin doublet, already expected to be rather weak, is not noticeable (no feature VI) at low magnetic fields ($<1$~T) in Fig. \ref{fig:1}(a), top row~\cite{footnote2}.

%To conclude, in this work we performed experimental and theoretical analysis of the transport spectra for a few-electron QDs. The experimental tunneling spectra reveal a number of the features (current steps) which cannot be explained solely by the QD energy spectrum. Rather, a full quantum mechanical treatment of the current through the QD is required to describe adequatly the experimental data from a theoretical point of view. By computing the current, we found that the assymetry in the tunneling barriers and spin selectivity of the current play the major role in assigning the observed features to the corresponding QD chemical potentials. On the other hand, quantum mechanical overlaps become important in description of the tunneling involving the excited states. In particular, features in the current plots due to the involvement of the excited states, the spin blocking of the transport channels, and due to the channels going out of (or in) the transport window are identified and the current changes at these features are analyzed. Overall, we found a good agreement between the experimental and computed tunneling spectra for both feature positions and tunneling current values.

We are grateful to R. M. Martin for helpful discussions. This work was supported by Grant-in-Aid for Scientific Research A, SORST-JST, IT program MEXT, DARPA QuIST, the MCC through the NSF, and the MRL through the USDE.


\begin{thebibliography}{100}
\bibitem{footnote1} The smaller computed magnetic field for the singlet-triplet crossing and the larger ground state addition energy (given by the separation between the $N=2$ and $N=3$ ground state curves) are due to the 2D parabolic approximation. For these quantities the full-scale 3D calculations are in a better agreement with the experimental data.
\bibitem{footnote2} In the vicinity of the emitter barrier, changes in the potential profile of the conduction band edge \cite{Matagne} due to the doping and the depletion of the leads may in principle affect the values of $\gamma_{L(R)}$, though in the QD studied this change is expected to be much smaller than the ratio $\gamma_L/\gamma_R\sim 7$.
\bibitem{footnote3} Note that in reverse bias the measured current along feature VIII actually slightly decreases rather than increases. The change in current is nonetheless much smaller than the corresponding decrease in Fig. \ref{fig:1}(a) so that the discrepancy can be easily caused by uncertainty in $\gamma$.
\end{thebibliography}
\end{document}